\documentclass[%
superscriptaddress,
showpacs,
preprint,
amsmath,amssymb,
aps,
floatfix,
dvipdfmx,
]{revtex4-1}

\usepackage{graphicx}
\usepackage{dcolumn}
\usepackage{bm}
\usepackage{multirow}
\usepackage[mathlines]{lineno}
\usepackage{color}
\usepackage{comment}
\usepackage[normalem]{ulem}

\begin{document}


\title{Two-Carrier Analyses of the Transport Properties of Black Phosphorus under Pressure}

\author{Kazuto~Akiba}
\email{k\_akiba@issp.u-tokyo.ac.jp}
\affiliation{
The Institute for Solid State Physics, University of Tokyo, Kashiwa, Chiba 277-8581, Japan
}

\author{Atsushi~Miyake}
\affiliation{
The Institute for Solid State Physics, University of Tokyo, Kashiwa, Chiba 277-8581, Japan
}

\author{Yuichi~Akahama}
\affiliation{
Graduate School of Material Science, University of Hyogo, Kamigori, Hyogo 678-1297, Japan
}

\author{Kazuyuki~Matsubayashi}
\affiliation{
The Institute for Solid State Physics, University of Tokyo, Kashiwa, Chiba 277-8581, Japan
}

\author{Yoshiya~Uwatoko}
\affiliation{
The Institute for Solid State Physics, University of Tokyo, Kashiwa, Chiba 277-8581, Japan
}

\author{Masashi~Tokunaga}
\affiliation{
The Institute for Solid State Physics, University of Tokyo, Kashiwa, Chiba 277-8581, Japan
}

\date{\today}


\begin{abstract}
We report on the electronic transport properties of black phosphorus and analyze them
using a two-carrier model in a wide range of pressure up to 2.5 GPa.
In semiconducting state at 0.29 GPa, the remarkable non-linear behavior in the Hall resistance is reasonably reproduced by assuming the
coexistence of two kinds of hole with different densities and mobilities. 
On the other hand, two-carrier analyses of the magnetotransport properties above 1.01 GPa suggest
the coexistence of high mobility electron and hole carriers that have
almost the same densities,
\textit{i.e.}, nearly compensated semimetallic nature of black phosphorus.
In the semimetallic state, analyses of both the two-carrier model and quantum oscillations
indicate a systematic increase in the carrier
densities as pressure increases.
An observed sign inversion of Hall resistivity at low magnetic fields suggests the existence of high mobility
electrons ($\sim$10$^5$ cm$^2$ V$^{-1}$ s$^{-1}$)
that is roughly ten times larger than that of holes, in the semimetallic black phosphorus.
We conclude that the extremely large positive magnetoresistance that has been observed in
semimetallic state cannot be reproduced by a conventional two-carrier model.
\end{abstract}

\maketitle

\section{introduction}
In condensed matter physics, one of the most important area of investigation is into
the transport properties of solids under magnetic fields.
The recent discovery of extremely large magnetoresistance (XMR) in non-magnetic materials
has attracted attention not only for interests in its physical mechanism but also its possible application 
in magnetic sensors.
Several non-magnetic semimetals, with nearly compensated and highly mobile
electrons and holes, have been known to exhibit XMR, such as WTe$_2$ \cite{Ali}, Cd$_3$As$_2$ \cite{Liang},
TaAs \cite{Huang}, and NbP \cite{Shekhar}.
Although several mechanisms, such as topologically protected back scattering \cite{Liang},
have been proposed to explain XMR, the exact origin still remains unclear.

Black phosphorus (BP), the target of this study, is one of the stable forms of phosphorus (P) and
known as a high-mobility semiconductor at ambient pressure.
BP has an orthorhombic crystal structure with distinctive layers in the $ac$ plane,
as shown in the lower image of Fig. \ref{fig1};
these layers consist of puckered honeycomb monolayers called phosphorene \cite{Hultgren}.
Phosphorene is recently shed light on the application to electronic devices owing to its high mobility and appropriate band gap ($\sim$2 eV) at the $\Gamma$ point \cite{Li}.
The interlayer coupling along the $b$-axis causes changes in the dispersion along the $\Gamma$-$Z$ path,
thereby resulting in bulk BP having a narrow direct gap ($\sim$0.3 eV) at the $Z$ point \cite{Takao, Asahina}. 
The band gap can be further reduced by applying hydrostatic pressure,
as the layers are weakly coupled by the van der Waals force.
Suppression of the band gap has been observed in transport and optical measurements under high pressure \cite{Okazima, Akahama, Akahama_1986}.
Recently, a pressure-induced semiconductor to semimetal (SC-SM) transition was directly observed
through the emergence of Shubnikov-de Haas (SdH) oscillations,
and the existence of at least two small Fermi surfaces are suggested in the semimetallic state \cite{Akiba, Xiang}.
In this state, BP showed giant 
and non-saturating magnetoresistance ($\sim$10$^5$ \% at 14 T) \cite{Akiba},
which is a feature frequently seen in nearly compensated semimetals \cite{Ali, Liang, Huang, Shekhar}.
By applying a conventional two-carrier model to compensated semimetals,
we can explain the effect of non-saturating positive transverse magnetoresistance
regardless of their topological properties.
Therefore, in this class of materials, careful two-carrier analyses are crucial in distinguishing
non-trivial effects from conventional ones.
In this context, we measured the magnetotransport properties of BP under hydrostatic pressure, and analyzed the results based on a two-carrier model.
 
\section{experiments}
Single crystals of BP were synthesized under high pressure \cite{Endo}.
The resistance ($\rho_{xx}$) and Hall resistance ($\rho_{yx}$) of samples that had average dimensions of
1.5 $\times$ 1.0 $\times$ 0.1 mm$^3$ were measured simultaneously by a standard five-probe method.
Gold wires of 30 $\mu$m diameter were attached to the samples with a MRX-713J carbon epoxy.
A high pressure environment was created by a piston-cylinder-type pressure cell,
and Daphne 7373 was used as a pressure medium.
The pressure ($P$) in the sample space was monitored by the superconducting-transition temperature of Pb, which had been mounted together with the sample.
The magnetotransport properties were studied in magnetic fields of up to 14 T and temperatures
above 2 K using the Physical Property Measurement System (PPMS, Quantum Design).
The measurements below 2 K were carried out using a dilution refrigerator and superconducting magnet
(Oxford Instruments).

\section{Results and Discussion}

\begin{figure}
\begin{center}
\includegraphics[width=7.5cm]{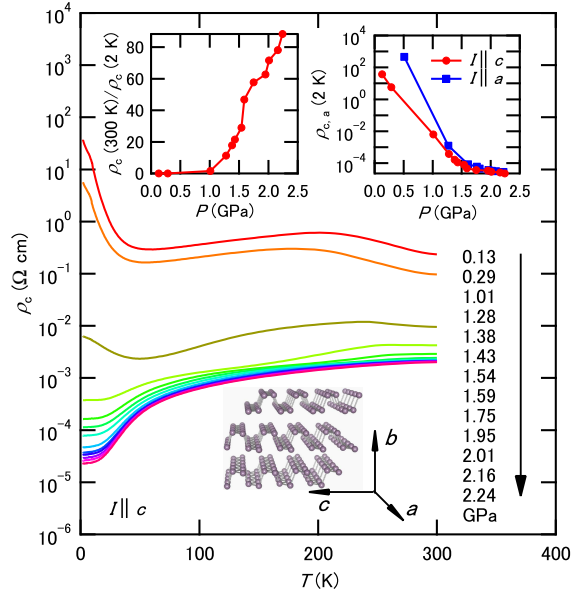}
\caption{The temperature dependence of the resistivity along the $c$-axis ($\rho_{c}$) at various pressures.
The upper left inset shows the pressure dependence of the resistivity ratio between 300 K and 2 K
[$\rho_{c}$(300 K) / $\rho_{c}$(2 K)].
{\color{red}{
The upper right inset shows the pressure dependence of the resistivity along the $c$- ($\rho_{c}$) and
$a$- ($\rho_{a}$) axes at 2 K.
}}
The lower image shows the crystal structure of BP and the corresponding crystal axes. \label{fig1}}
\end{center}
\end{figure}

Figure \ref{fig1} shows the temperature ($T$) dependence of the resistivity along the $c$-axis ($\rho_c$) at various pressures.
The behaviors at 0.13 and 0.29 GPa are typical of extrinsic semiconductors \cite{Akahama, Hummel}.
In the high-temperature region between 300 and 200 K, the resistivity increases
as the temperature decreases,
which can be regarded as a reduction in the number of carriers that are thermally activated
\textit{via} the excitation across the band gap.
In the middle-temperature region between 200 and 50 K,
the reduction of $\rho_{c}$ as the temperature decreases can be interpreted as an enhancement
in the mobility
caused by the suppression of lattice vibrations in this temperature region.
The steep increase in resistivity below 50 K can be attributed to both the suppression of thermally excited 
carriers from the impurity levels and Anderson localization \cite{Baba}.
Enhancements of $\rho_{c}$ below 50 K and above 200 K are suppressed as the pressure increases,
and the temperature dependence becomes metallic across the entire temperature range above 1.38 GPa.
The upper left inset in Fig. \ref{fig1} shows the pressure dependence of the resistivity ratio
between 300 K and 2 K [$\rho_{c}$(300 K) / $\rho_{c}$(2 K)].
The slope of the ratio changes at around 1.2 GPa, which is where 
the magnetoresistance starts to increase [as shown in Fig. \ref{fig3}(a)]; this is consistent with
the previous report \cite{Xiang}.

\begin{figure}
\begin{center}
\includegraphics[width=7.5cm]{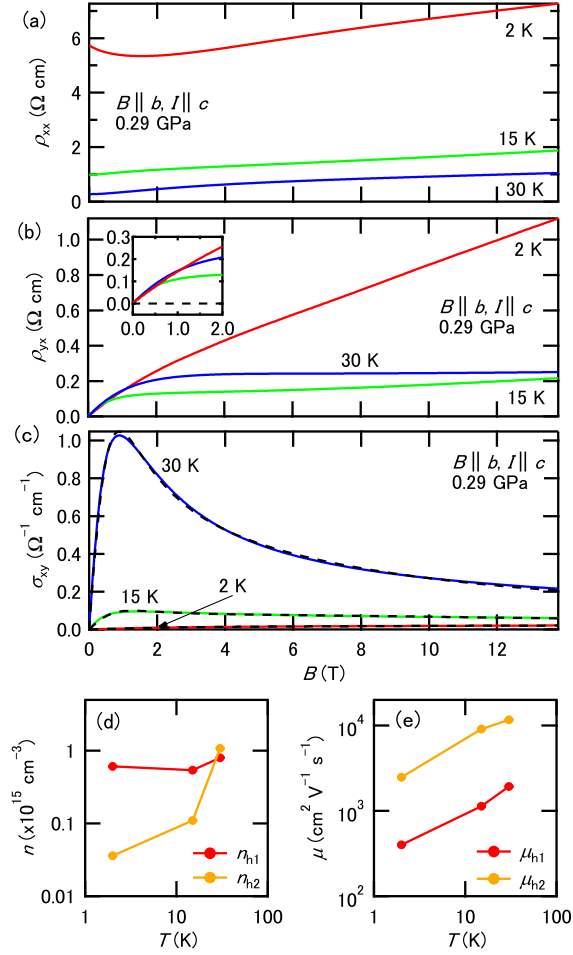}
\caption{
(a)The resistivity ($\rho_{xx}$) and (b)Hall resistivity ($\rho_{yx}$)
as a function of the magnetic field that is applied along the $b$-axis at 2, 15, and 30 K
under $P=0.29$ GPa.
The inset of (b) shows a magnified view of the weak magnetic field region below 2 T.
(c)The Hall conductivity ($\sigma_{xy}$) at 2, 15, and 30 K, as
calculated from both $\rho_{xx}$ and $\rho_{yx}$ using Eq. (\ref{eq_sigmaxy}).
Dashed lines indicate the fitted curves based on the two-carrier model (see text).
The temperature dependence of the (d)densities ($n_{h1}$, $n_{h2}$) and (e)mobilities ($\mu_{h1}$, $\mu_{h2}$)
of the holes estimated by the two-carrier model fittings at 0.29 GPa.
The error bars are smaller than the diameter of the circles.
\label{fig2}}
\end{center}
\end{figure}

The transport properties in the semiconducting state at 0.29 GPa are worth particular attention.
Figure \ref{fig2}(a) shows $\rho_{xx}$ at $T=2$, 15, and 30 K as a function of the magnetic field
that is applied parallel to the $b$-axis.
$\rho_{xx}$ first decreases below 2 T and then increases as the magnetic field increases at 2 K;
it then monotonically increases with the magnetic field at 15 and 30 K.
Negative magnetoresistance below 2 T at 2 K is ascribed to the
suppression of the two-dimensional Anderson localization \cite{Baba, Strutz}.
Figure \ref{fig2}(b) shows $\rho_{yx}$ at 2, 15, and 30 K.
Non-linear behavior below 2 T becomes pronounced as the temperature
increases, as shown in the inset of Fig. \ref{fig2}(b);
this agrees well with the previous report \cite{Hou}.
This non-linear behavior suggests that multiple kinds of carriers exist,
and that they have different densities and mobilities.
Since $\rho_{yx}$ is always positive from 0 T to at least 14 T,
the majority of the carriers can be considered to be holes.
Figure \ref{fig2}(c) shows the
Hall conductivity ($\sigma_{xy}$)
as calculated from both $\rho_{xx}$ and $\rho_{yx}$.
$\sigma_{xy}$ is defined by $\rho_{xx}$ and $\rho_{yx}$:
\begin{equation}
\sigma_{xy}=\frac{\rho_{yx}}{\rho_{xx}^2+\rho_{yx}^2}.
\label{eq_sigmaxy}
\end{equation}
{\color{red}{
Here, Eq. (\ref{eq_sigmaxy}) is for the isotropic case with $\rho_{xx}=\rho_{yy}$
and $\rho_{xy}=-\rho_{yx}$ in the original definition:
\begin{equation}
\sigma_{xy}=\frac{\rho_{yx}}{\rho_{xx}\rho_{yy}-\rho_{xy}\rho_{yx}}.
\end{equation}
In the case of BP, It has been reported that the resistivity along the $a$-axis ($\rho_{a}$) is approximately
10 times larger than $\rho_{c}$ at ambient pressure \cite{Akahama_Hall}.
We actually observed the anisotropy in the semiconducting state as shown in
the upper right inset of Fig. \ref{fig1}.
Since $\rho_a/\rho_c$ was found to be nearly constant as functions of temperature and
magnetic field, we utilized the isotropic two-carrier model and focus on qualitative change in the
parameters in the semiconducting state.
The upper right inset of Fig. \ref{fig1} also shows that BP becomes nearly isotropic at pressures above 1.2 GPa.
Therefore, isotropic two-carrier analyses becomes more accurate in the semimetallic state.
}}
By assuming two kinds of carriers (labeled 1 and 2) with densities ($n_{1,2}$) and mobilities ($\mu_{1,2}$),
we can obtain $\sigma_{xy}$ \cite{Soule}:
\begin{equation}
\sigma_{xy}=eB\left(\pm \frac{n_1\mu_1^2}{1+\mu_1^2B^2}\pm \frac{n_2\mu_2^2}{1+\mu_2^2B^2} \right).
\label{eq_sigmaxy_ele_hole}
\end{equation}
In this equation, $e>0$ is the elementary charge and $B$ is the magnetic flux density, respectively.
The plus and minus signs in Eq. (\ref{eq_sigmaxy_ele_hole}) should be taken into consideration
when the corresponding carrier is a hole or electron, respectively.
It should be noted that the signs of $\rho_{yx}$ and $\sigma_{xy}$ are the same in this formulation.
Our discussion of the result in the semiconducting state assumes the coexistence of two kinds of hole
carries [\textit{i.e.}, plus signs are taken in Eq. (\ref{eq_sigmaxy_ele_hole})].
The dashed lines in Fig. \ref{fig2}(c) show the fitted curves of $\sigma_{xy}$,
based on Eq. (\ref{eq_sigmaxy_ele_hole}), and we
adjust for the four fitting coefficients; namely,
the densities ($n_{h1,h2}$) and mobilities ($\mu_{h1,h2}$) of the hole carriers.
The temperature dependence of these coefficients is shown in Figs. \ref{fig2}(d) and (e).
For simplicity, we ignore the effects of the Anderson localization that are observed at 2 K.
We can see that the carrier densities ($n_{h1,h2}$) increase with increasing temperature, which can be explained \textit{via} the
thermal excitation of carriers from the impurity levels.
The mobilities ($\mu_{h1}$ and $\mu_{h2}$) also increase with increasing temperature,
which is consistent with the
positive temperature coefficient of the Hall mobility that had been observed at ambient pressure up to 20 K
 \cite{Akahama_Hall}.

Here, we comment on the statistical error accompanying the two-carrier fit.
Eguchi \textit{et al.} recently pointed out that there is a large uncertainty in the fitting coefficients
of the two-carrier model analyses of the resistivities ($\rho_{xx}$ and $\rho_{yx}$) \cite{Eguchi_PRB, Eguchi}.
We also recognized that there are significant errors in the fitting coefficients in the analyses of resistivities,
and so we adopted a fitting to the Hall conductivity ($\sigma_{xy}$).
The parameters we obtained had a much lower error and resulted in physically reasonable behavior, as mentioned above.
One important point to make is that, contrary to that of previous report on BP \cite{Hou},
our experimental results cannot be reproduced by
assuming the coexistence of electrons and holes;
our results indicate that the existence of two kinds of hole carriers with different densities and mobilities is crucial to reproduce the experimental results.

\begin{figure}
\begin{center}
\includegraphics[width=7.5cm]{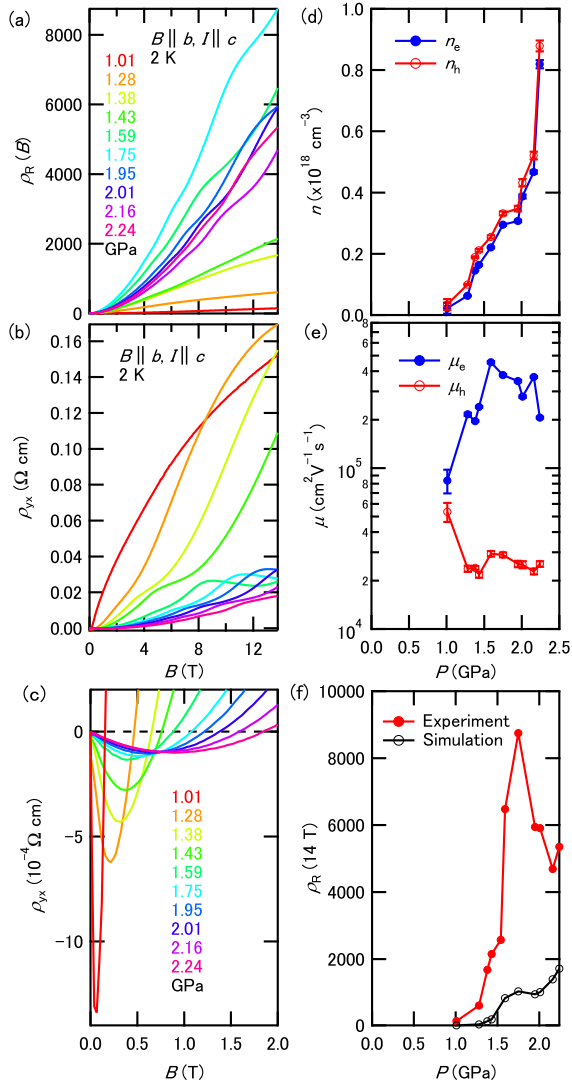}
\caption{
(a)The pressure dependence of $\rho_{\mathrm{R}}(B)=[\rho_{xx}(B)-\rho_{xx}(0$ T$)]/\rho_{xx}(0$ T$)$
from 1.01 to 2.24 GPa at 2 K.
(b)The pressure dependence of the Hall resistivity ($\rho_{yx}$) at 2 K.
(c)The magnified view of $\rho_{yx}$ below 2 T.
(d)The pressure dependence of the hole ($n_h$) and electron ($n_e$) densities with vertical error bars.
(e)The pressure dependence of the hole ($\mu_h$) and electron ($\mu_e$) mobilities with vertical error bars.
(f)The experimental (filled red circles) and
simulated (open black circles) values of $\rho_{\mathrm{R}}(14$ T$)$.
\label{fig3}}
\end{center}
\end{figure}

Next, we show the magnetoresistance and Hall resistance near and above the SC-SM transition pressure.
Figure \ref{fig3}(a) shows $\rho_R (B) \equiv [\rho_{xx }(B)-\rho_{xx} (0$ T$)]/\rho_{xx} (0$ T$)$ at 2 K at several pressures from 1.01 to 2.24 GPa.
The details about the SdH oscillation superposed on $\rho_R (B)$, meanwhile, will be discussed later.
We do not see any tendency of saturation in $\rho_R (B)$ until at least 14 T in the semimetallic state.
The magnetoresistance effect reaches its maximum value of $\rho_R (14$ T$)\cong 8000$ at 1.75 GPa and then decreases with increasing pressure, as shown in Fig. \ref{fig3}(f);
meanwhile the $\rho_{c}$ at 2 K is monotonically suppressed by pressure,
as shown in Fig. \ref{fig1}.
Figure \ref{fig3}(b) shows $\rho_{yx}$ at 2 K from 1.01 to 2.24 GPa, and it
shows non-linear behavior at all pressures,
which suggests a contribution from multiple kinds of carriers.
The sign inversions of $\rho_{yx}$ were observed at this pressure region below 2 T,
as shown in Fig. \ref{fig3}(c), which agrees with the previous report \cite{Xiang};
this sign inversion in $\rho_{yx}$, however, cannot be recognized at 0.29 GPa, as shown in
the inset of Fig. \ref{fig2}(b). 
The magnetic field where the sign inversion takes place systematically increases with increasing pressure.

In order to extract quantitative information about the semimetallic BP, we analyzed $\sigma_{xy}$ at 2 K based on the isotropic two-carrier model.
{\color{red}{
As shown in the upper right inset of Fig. \ref{fig1}, the anisotropy between $\rho_{a}$ and $\rho_{c}$
at 2 K becomes small above  1.2 GPa at zero field.
In addition, the magnetoresistance along the $a$- and $c$- axes was reported to be less anisotropic
in the semimetallic state \cite{Xiang}.
}}
We can reasonably reproduce $\sigma_{xy}$ by assuming the coexistence of electron and hole carriers
at pressures from 1.01 to 2.24 GPa
[opposite signs between the first and second terms are taken in Eq. (\ref{eq_sigmaxy_ele_hole})].
Curve fittings were carried out in the magnetic field from 0 T to 4 T where the effect of the
quantum oscillations is negligible.
Figure \ref{fig3}(d) shows the pressure dependence of the electron ($n_e$) and hole ($n_h$) densities
with vertical error bars;
both $n_e$ and $n_h$ monotonically increase with increasing pressure,
which indicates that the Fermi pockets became larger due to the enhancement of the band overlap.
This result shows that carrier density is continuously controlled by hydrostatic pressure
in semimetallic BP.
It is also shown in Fig. \ref{fig3}(d) that $n_{e}$ and $n_{h}$ have similar values at all pressures,
which indicates the nearly compensated semimetallic nature of BP.
Figure \ref{fig3}(e) shows the pressure dependence of the electron ($\mu_e$) and hole ($\mu_h$) mobilities
with vertical error bars;
The $\mu_{e}$ and $\mu_{h}$ have similar values at 1.01 GPa, while $\mu_e$ becomes
more than 10 times larger than $\mu_h$ above 1.28 GPa.
The coexistence of electrons and holes,  and the large difference between $\mu_{e}$ and $\mu_{h}$, reasonably explains the sign inversion of
$\rho_{yx}$ that is shown in Fig. \ref{fig3}(c).
In the electron-hole two-carrier model, $\sigma_{xy}$ is represented by Eq. (\ref{eq_sigmaxy_ele_hole}):
\begin{equation}
\sigma_{xy}=eB\left( \frac{n_h}{\mu_h^{-2}+B^2}-\frac{n_e}{\mu_e^{-2}+B^2} \right).
\label{eq_sigxy}
\end{equation}
In the strong-magnetic field limit ($\mu_{e,h}^{-1} \ll B$), Eq. (\ref{eq_sigxy}) can be simplified:
\begin{equation}
\sigma_{xy}\sim \frac{e}{B}(n_h-n_e).
\label{eq_sigxy_high}
\end{equation}
This equation indicates that the sign of $\sigma_{xy}$ is only determined by the carrier imbalance,
$n_h-n_e$.
$\sigma_{xy}$ is always positive in the strong-field limit in the present case,
since $n_h > n_e$ holds at all pressures.
In the weak-magnetic field limit ($\mu_{e,h}^{-1} \gg B$), on the other hand, $\sigma_{xy}$ is represented
by the following equation:
\begin{equation}
\sigma_{xy}\sim eB(\mu_h^2 n_h -\mu_e^2 n_e).
\label{eq_sigxy_weak}
\end{equation}
As can be seen, the sign depends also on the mobilities. Since $n_e \simeq n_h$ and $\mu_e^2 > \mu_h^2$ in the present case, $\sigma_{xy}$ is negative in a weak magnetic field.
Therefore, the sign inversion takes place in a magnetic field where the first term is equal to the second term in Eq. (\ref{eq_sigxy}).

Here, we discuss the XMR of semimetallic BP.
In the present model, $\rho_{\mathrm{R}}(B)$ is represented by the following equation \cite{Sun}:
\begin{equation}
\rho_{\mathrm{R}}(B)=\frac{B^2n_e n_h \mu_e \mu_h (\mu_e+\mu_h)^2}{(\mu_e n_e+\mu_h n_h)^2+B^2\mu_e^2 \mu_h^2(n_e-n_h)^2}.
\label{eq_rhoR}
\end{equation}
In the case of a completely compensated semimetal ($n_e=n_h$), $\rho_{R}(B)$ can be reduced to $\mu_e \mu_h B^2$;
the resistance continuously increases without showing saturation within the classical model.
In the present case, however, incomplete compensation leads to the saturation of $\rho_{R}(B)$ to a finite value.
We calculated $\rho_{R}(14$ T$)$ by Eq. (\ref{eq_rhoR}) using $n_{e,h}$ and $\mu_{e, h}$, as shown in Figs. \ref{fig3}(d) and (e).
The open circles in Fig. \ref{fig3}(f) show that the calculated $\rho_{R}(14$ T$)$
is smaller than the experimental values at all pressures, and it
cannot reproduce the peak structure at 1.75 GPa.
Therefore, the observed XMR in BP involves some additional physics beyond the conventional two-carrier model for nearly compensated semimetals.
Additionally, the observed large longitudinal magnetoresistance, which is shown later,
cannot be explained through this semi-classical approach.

\begin{figure}
\begin{center}
\includegraphics[width=7.5cm]{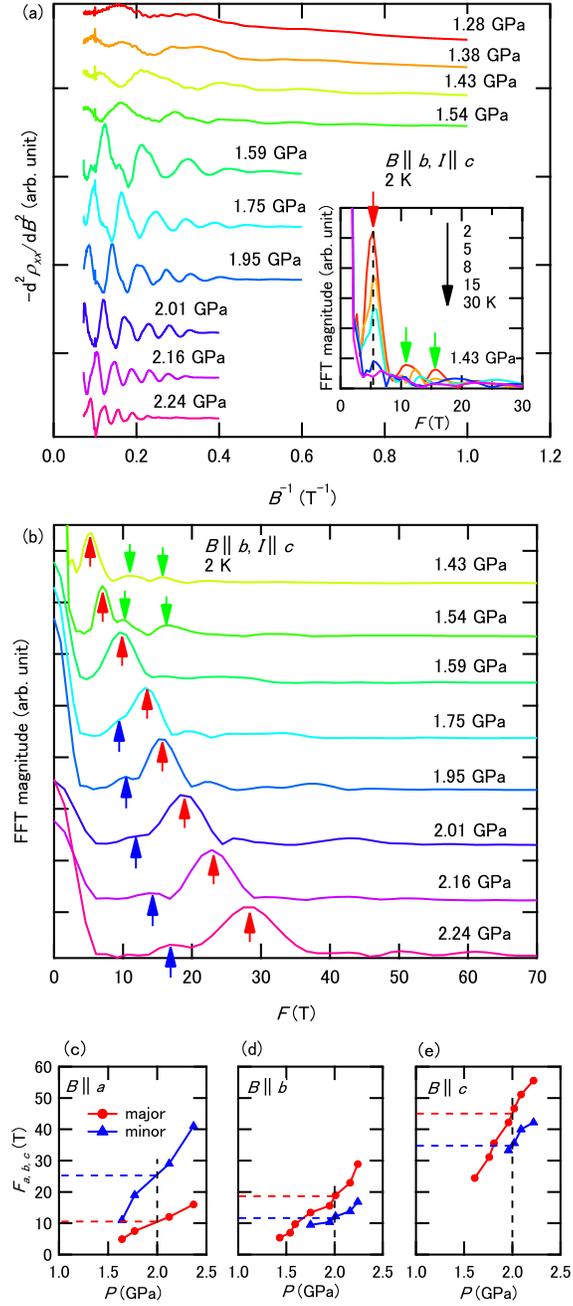}
\caption{
{\color{red}{(a)The pressure dependence of $-d^2\rho_{xx}/dB^2$ at 2 K in magnetic fields along the $b$-axis. The data were vertically offset for clarity.
The inset shows the temperature dependence of the FFT spectra at 1.43 GPa.
(b)The pressure dependence of the FFT spectra of $-d^2\rho_{xx}/dB^2$ from 1.43 to 2.24 GPa at 2 K.
The major and minor peaks are indicated by the red and blue arrows, respectively.
Several additional peaks which appear at 1.43 and 1.54 GPa are indicated by the green arrows.
Each magnitude of the spectrum is normalized by the amplitude of each major peak
and vertically offset for clarity.
The pressure dependence of the peak frequencies in magnetic fields along the (c)$a$-, (d)$b$-, and (e)$c$-axes. In all of the field directions, the major peak with a large FFT amplitude (red),
and the minor peak with smaller one (blue), were observed.}}   
\label{fig4}}
\end{center}
\end{figure}

\begin{figure}
\begin{center}
\includegraphics[width=7.5cm]{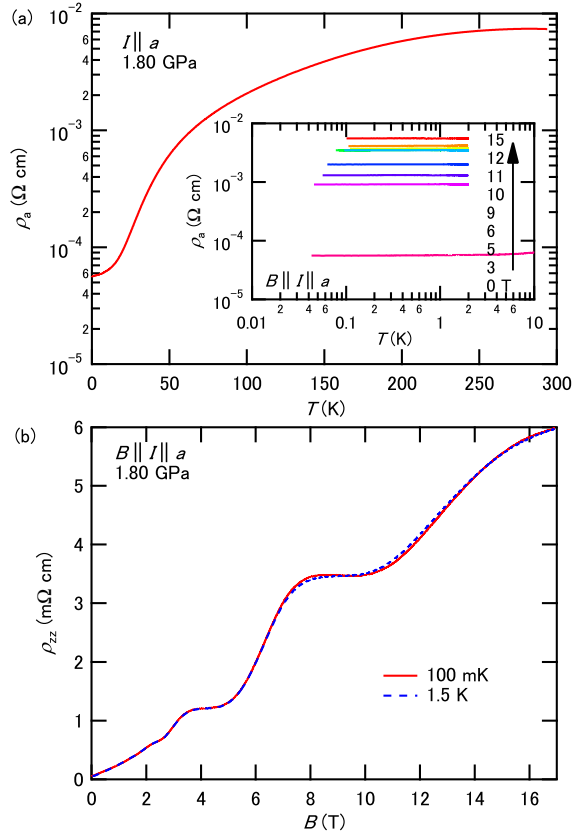}
\caption{
(a)The temperature dependence of the resistivity along the $a$-axis ($\rho_{a}$) at 1.80 GPa down to 43 mK. 
The inset shows the temperature dependence of $\rho_{a}$ below 1 K under various magnetic fields along
the $a$-axis.
(b)The magnetoresistivity ($\rho_{zz}$) at 100 mK and 1.5 K in the longitudinal configuration
($B \parallel I \parallel a$).
\label{fig5}}
\end{center}
\end{figure}

{\color{red}{
Now, we focus on the SdH oscillations and the Fermi surfaces of the semimetallic BP.
Figure \ref{fig4}(a) shows the pressure dependence of $-d^2\rho_{xx}/dB^2$ at 2 K in magnetic fields
applied along the $b$-axis.
As the pressure increases, oscillatory structure becomes more prominent, and the frequency clearly
becomes larger.
Figure \ref{fig4}(b) shows the pressure dependence of the fast Fourier transform (FFT) spectra from 1.43 to 2.24 GPa at 2 K.
FFT was carried out for $-d^2\rho_{xx}/dB^2$ shown in Fig. \ref{fig4}(a) with Hanning window function.
We identified two frequency peaks marked with red (referred to as a major peak)
and blue (referred to as a minor peak) arrows.
The Major peak first becomes discernible at 1.43 GPa with a frequency of about 5 T,
while the minor peak appears at 1.75 GPa with a frequency of about 9.5 T.
Both frequencies become large as pressure increases, which is consistent with the pressure dependence of the carrier densities estimated from the two-carrier analyses [Fig. \ref{fig3}(d)].
Although SdH oscillation-like structures can be seen in $-d^2\rho_{xx}/dB^2$
at 1.28 and 1.38 GPa, we cannot define the reliable frequency due to the limited number of cycles of the oscillations.
We also note that there are some additional structures at higher frequencies than that of major peaks
as marked by green arrows at 1.43 and 1.54 GPa in Fig. \ref{fig4}(b).
Since these peaks do not show systematic dependence on temperature as shown in the inset of Fig. \ref{fig4}(a), we focus on the other two peaks in the present discussion.
In all of the orientations,
two branches, with larger and smaller amplitudes are observed
as shown in Figs. \ref{fig4}(c) to (e);
they will be also referred to as the major and minor peaks, respectively. 
We can estimate the SC-SM transition pressure as 1.2-1.4 GPa where FFT frequencies for the magnetic field along the three principal axes ($F_{a-c}$) becomes zero
by extrapolating from the data shown in Figs. \ref{fig4}(c)-(e).
Since we did not find any other peaks down to the lowest temperatures in this study [as shown in
Fig. \ref{fig5}(b)],
we attribute these two frequencies to that of the quantum oscillations from the electron and hole
Fermi pockets.
We cannot, however, identify which frequency corresponds to the electron/hole pocket for
all of the three field directions.
}}

The first-principles calculation from a previous report predicted that semimetallic BP has
one anisotropic biconcave hole pocket at the $Z$ point, and four relatively isotropic electron pockets on the $\Gamma$-A path \cite{Akiba}.
For the nearly compensated condition, the total volume of the electron pockets should be equal to
that of the hole pocket.
For the sake of simplicity, we approximate the Fermi surfaces as spheroids that are characterized by three axes, $L_a$, $L_b$, and $L_c$; \textit{i.e.}, the Fermi surfaces are represented in the reciprocal space by
$(k_a/L_a)^2+(k_b/L_b)^2+(k_c/L_c)^2=1$.
Here the $a$-, $b$-, and $c$-directions in the real space correspond to the $k_a$-, $k_b$-, and $k_c$-
directions in the reciprocal space. 
From the first-principles calculation, the anisotropy of the hole pocket is estimated as $L_a:L_b:L_c=21:13:3$ at 2 GPa \cite{Akiba}; as a result, the ratio of the cross-sections perpendicular to the $a$- ($S_a$), $b$- ($S_b$), and $c$- ($S_c$) axis is $S_a:S_b:S_c=13:21:91$.
By assuming that the major peaks originated from the hole pockets
{\color{red}{and using extrapolated $F_{a, b, c}$ at 2 GPa
[indicated with dashed lines in Fig. \ref{fig4}(c)-(e)],}} 
the hole density calculated from the volume of the spheroid is about $2.6\times 10^{17}$ cm$^{-3}$,
which is the same order of magnitude as that derived from the two-carrier analysis shown in Fig. \ref{fig3}(d).
In addition, this identification results in the relationship $S_c > S_b > S_a$,
which was suggested by the theoretical
calculation.
In this case, however, the electron pocket has an almost identical volume to that of the hole,
even for one pocket,
which does not match the first-principles calculation.
{\color{red}{Recently, several theoretical studies reported different band structures of BP in the semimetallic state,
which also could not reproduce the observed SdH frequencies in the present study \cite{Gong, Zhao}.
Additional information, such as an angle-resolved SdH measurements and its careful comparison
with theoretical studies, are necessary for determining
the structures of the Fermi surfaces in BP. 
}}

Finally, we focus on the magnetotransport properties at lower temperatures,
which was achieved through the use of a dilution
refrigerator.
Since semimetallic BP is an ideally clean low-carrier system,
shielding of the Coulomb interaction will be weaker than in ordinary metals.
In order to look for anomalous quantum states caused by the charge correlation in this electron-hole system,
we studied the magnetoresistance of BP at low temperatures.
Figure \ref{fig5}(a) shows the temperature dependence of $\rho_{a}$ at $P=1.80$ GPa in various fields
that were
applied along the $a$-axis.
No anomalous features in the $\rho_{a}$-$T$ curves were observed down to 43 mK.

Figure \ref{fig5}(b) shows longitudinal magnetoresistance ($\rho_{zz}$) at $T=100$ mK.
The observed profile is almost identical to that observed at $T=1.5$ K.
No additional components in the oscillation can be identified down to this temperature.
High field studies on clean elemental semimetals of bismuth and graphite show anomalous behavior in the
vicinity of the quantum limit state
{\color{red}{\cite{Brandt, Kuchler, Tanuma, Iye_1982, Yaguchi_PRL, Fauque, Akiba_graphite}}}.
In the case of semimetallic BP, we do not find any features indicating a phase transition up to 17 T.
This result shows XMR with
{\color{red}{$\rho_{R}(17$ T$)=100$}}
even in the longitudinal configuration.
According to a conventional Drude-type formulation, $\sigma_{zz}(=1/\rho_{zz})$ is insensitive to $B$ and
described by the following equation:
\begin{equation}
\sigma_{zz}=\dfrac{n_h e^2 \tau_h}{m_h}+\dfrac{n_e e^2 \tau_e}{m_e}.
\end{equation} 
In this equation, $\tau_{h(e)}$ and $m_{h(e)}$ represent the relaxation time and the effective mass
of the holes (electrons), respectively.
Since the observed SdH oscillations can be analyzed using fixed values of $n_{h, e}$ and $m_{h, e}$,
we have to introduce a field dependence of $\tau_{h, e}$ so as
to reproduce the observed longitudinal magnetoresistance in this classical framework.

\section{conclusion}
We investigated the transport properties of black phosphorus under pressure.
In the semiconducting state at 0.29 GPa, the existence of two kinds of hole carriers that have different
densities and mobilities is crucial to reproduce the experimental results.
Above 1.01 GPa, the electron-hole two-carrier model consistently explains the transport properties.
In the semimetallic state above 1.43 GPa, it is quantitatively shown that approximately the same number of electrons and holes contribute to the transport, and that their densities become larger as pressure
is increased.
Additionally, the mobility of the electrons is about ten times larger than that of the holes,
which explains the sign inversion seen for $\rho_{yx}$.
The analyses made through a conventional two-carrier model were found to be insufficient in explaining
the extremely large magnetoresistance observed in semimetallic black phosphorus under high pressure.

\begin{acknowledgments}
We thank Y. Fuseya for valuable discussion and comments
{\color{red}{and L. Zou for showing us in detail results of band calculations}}.
This work was supported by JSPS KAKENHI Grant Number 15K17700.
K. A. was supported by Grant-in-Aid for JSPS Research Fellow (16J04781).
\end{acknowledgments}

\bibliography{ref}

\end{document}